\documentclass[sigconf]{acmart}
\AtEndOfPackage{\RequirePackage{hyperxmp}}
\usepackage{tabularx}
\usepackage{makecell}

\AtBeginDocument{%
  \providecommand\BibTeX{{%
    \normalfont B\kern-0.5em{\scshape i\kern-0.25em b}\kern-0.8em\TeX}}}

\copyrightyear{2025}
\acmYear{2025}
\setcopyright{acmlicensed}
\acmConference[CHI '25]{CHI Conference on Human Factors in Computing Systems}{April 26-May 1, 2025}{Yokohama, Japan}
\acmBooktitle{CHI Conference on Human Factors in Computing Systems (CHI '25), April 26-May 1, 2025, Yokohama, Japan}
\acmDOI{10.1145/3706598.3713633}
\acmISBN{979-8-4007-1394-1/25/04}

\begin{document}

\title{Ontologies in Design: How Imagining a Tree Reveals Possibilites and Assumptions in Large Language Models}


\author{Nava Haghighi}
\email{nava@cs.stanford.edu}
\affiliation{%
  \institution{Stanford University}
  \streetaddress{450 Serra Mall}
  \city{Stanford}
  \state{CA}
  \country{USA}
  \postcode{94305}
}

\author{Sunny Yu}
\affiliation{%
 \institution{Stanford University}
 \city{Stanford}
 \state{CA}
 \country{USA}}
\email{syu03@stanford.edu}

\author{James Landay}
\affiliation{%
 \institution{Stanford University}
 \city{Stanford}
 \state{CA}
 \country{USA}}
\email{landay@stanford.edu}

\author{Daniela Rosner}
\affiliation{%
 \institution{University of Washington}
 \city{Seattle}
 \state{WA}
 \country{USA}}
\email{dkrosner@uw.edu}

\renewcommand{\shortauthors}{Haghighi, et al.}
\renewcommand{\shorttitle}{Ontologies in Design}

\begin{abstract}
Amid the recent uptake of Generative AI, sociotechnical scholars and critics have traced a multitude of resulting harms, with analyses largely focused on values and axiology (e.g., bias). While value-based analyses are crucial, we argue that ontologies---concerning what we allow ourselves to think or talk about---is a vital but under-recognized dimension in analyzing these systems. 
Proposing a need for a practice-based engagement with ontologies, we offer four orientations for considering ontologies in design: pluralism, groundedness, liveliness, and enactment. We share examples of potentialities that are opened up through these orientations across the entire LLM development pipeline by conducting two ontological analyses: examining the responses of four LLM-based chatbots in a prompting exercise, and analyzing the architecture of an LLM-based agent simulation. We conclude by sharing opportunities and limitations of working with ontologies in the design and development of sociotechnical systems.
\end{abstract}




\begin{teaserfigure}
        \centering
        \includegraphics[width=1\textwidth]{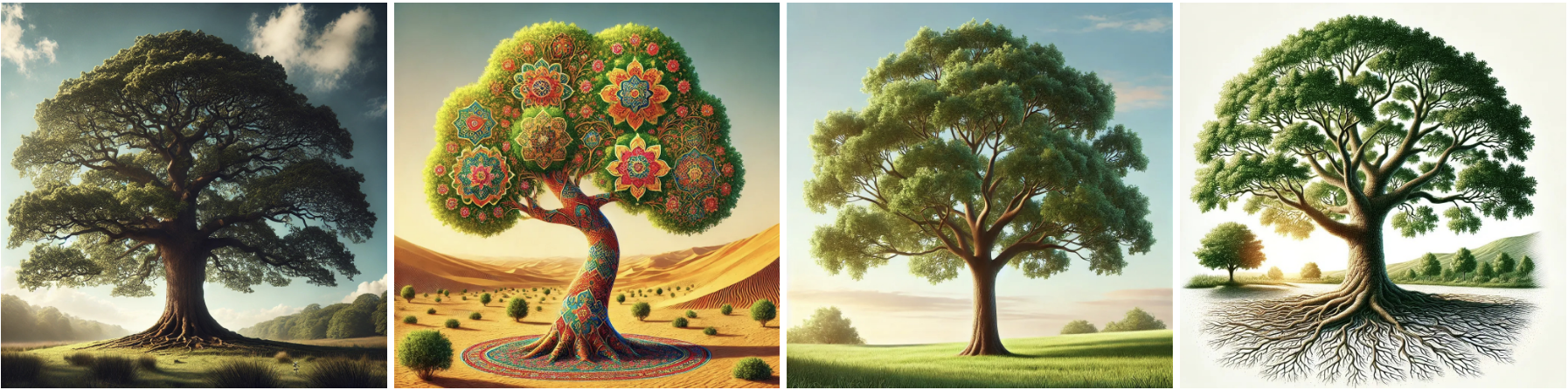}
        \caption{Images generated with GPT-4o. Prompts from left: (a) make me a picture of a tree. (b) I'm from Iran. make me a picture of a tree. (c) make me a picture of a tree [in Persian]. (d) everything in the world is connected. make me a picture of a tree.}
        \Description{To ensure the description is an accurate reflection of what GPT-4o intended to visualize, descriptions are verbatim the copy of the caption that was generated with the image. From left to right:
(a) A large, majestic tree with a thick trunk and sprawling branches. The tree is standing alone in a grassy field under a bright blue sky with scattered clouds.
(b) A vibrant Persian-inspired tree, standing in a desert-like landscape, with intricate patterns on the trunk and branches. The tree is full of life. [author note: this tree has colorful motifs resembling those seen in persian architecture. The motifs are cartoonish, not resembling any tree that can exist in real-life. They cover the entire tree from the leaves to the trunk and continuing around the base of the tree in a rug-like form]
(c) A beautiful, tall tree with a sturdy trunk and sprawling branches covered in green leaves, standing in a peaceful, grassy field under a clear blue sky.
(d) A detailed depiction of a single large tree standing tall in a natural environment. The tree has a sturdy, textured trunk and expansive branches. [author note: there are at least 10 other trees seen in the distance. This tree is the only tree with roots visualized even under ground]}
        \label{fig:trees}
\end{teaserfigure}

\received{12 September 2024}
\received[revised]{10 December 2024}
\received[accepted]{16 January 2025}

\maketitle

\section{Introduction}

When you think of a tree, what do you imagine? What does your tree feel like? Where have you encountered this tree before? What landscape does it exist in? How would \textit{you} describe it? The answers to these questions depend on how you have come to know a tree. If you are a botanist, you might imagine the mineral nutrients it accepts from neighboring fungi seeking carbohydrates. If you are a nutritionist, you might recall the mast-fruiting varieties that hold calories as starch in their roots. Or if you are a spiritual healer, you might picture trees whispering to one another, how they accept the delivery of mineral nutrients not as neatened biological processes but as gifts, taking care of us as we care for them. In her celebrated book \textit{Braiding Sweetgrass} \cite{kimmerer2013braiding}, author Robin Kimmerer, a moss ecologist and a member of the Citizen Potawatomi Nation, describes trees across each of these epistemic registers. For her, they are not opposing entities but rather interwoven with cosmologies that orient us to the world. They belong to ways of looking, feeling, and noticing that produce a source of identity and knowledge (\cite{kimmerer2013braiding}, p.24); an \textit{ontological} encounter. 

Ontology, the study of the nature of being, has long been core to humanistic lines of inquiry, and later to computing and information fields across the late 1980s and early 1990s to explain representations of data \cite{winograd1986understanding}. Yet ontology is notoriously difficult to comprehend. As historian of science Ian Hacking opines, ``If, like myself, you can understand the aims of psychology, cosmology, and theology, but are hard pressed to explain what a study of being in general would be, you can hardly welcome talk of ontology'' (\cite{hacking}, p.1). Hacking usefully focuses on ontology as a means of grappling with what we ``allow ourselves to talk about'' (\cite{hacking}, p.2).
Similarly, ethnographer and anthropologist Annemarie Mol articulates ontology in terms of potentialities or ``what belongs to the real, or conditions of possibility we live with''~\cite{mol1999ontological}. Mol looks to ``ontological politics'' to highlight the  ``process of shaping'' reality within those conditions of possibility ~\cite{mol1999ontological}. 

This notion of what we enable ourselves, as HCI analysts, to talk about or deem possible is essential to any inquiry into design formations, particularly in a moment of increasing attention to generative AI (Gen AI).
Many commentators have claimed that Gen AI has taken the world by surprise with its ability to complete many tasks that previously required human efforts \cite{Rothman_2024}. Critics have warned against these automated effects, identifying the differential experiences produced by Gen AI, with benefits and harms falling along existing lines of inequity such as race, gender, and disability \cite{stereotypesbianchi, zhang2024genderalign, nadeem2020stereoset, gadiraju2023wouldn}.
Within this body of work, efforts at algorithmic alignment, a process of ensuring that AI systems complement human goals, have largely focused on axiological concerns related to values and ethics~\cite{liu2022aligning, abbo2023social, kasirzadeh2023conversation}.
This focus on values entails a consideration of judgment around emphasis or selection in design decisions such as, returning to our tree example, the choice of landscape the tree produces, or the style of the leaves rendered. It also emphasizes selection absence \cite{sherman2024power}: the missing geographic locations (e.g., Iran) that an entity such as a tree might exist within; or the ignored pictorial techniques such as Souzandouzi, a needlework technique used in Iran.

While powerful and productive on their own, such values-oriented analyses may overlook questions of representation underlying the range of imagined possibilities: the underlying ontological assumptions. For example, what are the boundaries of a tree? Does the tree have distinct parts or is it all connected? Is it pictured with its roots? Is the tree imagined by its spiritual or symbolic significance or aesthetic dimensions? Is it a biological individual or part of a network of life such as a mycorrhizal network? Or as Kimmerer might ask, what does its imbued spirit have to say? 
Each of these questions has implications for not only what phenomena we observe and comprehend as \textit{belonging} to a category (demarcations), but also for what phenomena we imagine as even possible (realms of possibility). Take the example of how this analysis may play out in therapy applications, an increasingly popular area for Gen AI research. An axiological orientation may identify value tensions and harms such as identities that are implicitly ignored or explicitly censored \cite{reichenpfader-denecke-2024-simulating}. However, an ontological orientation may reveal additional considerations such as the possibility of a communal (compared with individualistic) healing or of connecting with spiritual traditions. Or if we envision a tree as limited to the parts visible above ground (trunk, branches, leaves, etc.), we may miss how its roots form interconnections with other trees as well as other species such as mycorrhizal fungi, creating life-sustaining symbiotic networks that exchange resources like water, carbon, and nitrogen. To picture these networks of exchange as somehow important or even sacred aspects of identifying a tree is to sensitize ourselves to the workings of the surrounding environment. It is also to push our imagination toward changes that challenge existing paradigms or propose transformative alternatives.

To be sure, an ontological shift expressed by a Gen AI prompt (e.g., create a therapy protocol, or visualize a tree network) will not solve major mental health or climate issues. But, as users of these systems, we may find that an alternative ontological approach has important albeit subtle consequences for our perspectives on our own social and ecological lives, shaping what we notice and concern ourselves with in connection with a phenomenon as common as a tree. As for Figure \ref{fig:trees}, ontological considerations can shift what comes into view. Despite iteratively grounding the prompt to ``visualize a tree''---representing what the model deems to be an ``Iranian'' aesthetic, for instance---the resulting outputs remain eerily similar. Each depicts a semicircle of symmetrically outstretched branches connected to an isolated trunk without visible roots or ecological connections.
A focus on ontology prompts HCI analysts to ask about what is taken-for-granted and what might be otherwise: \textit{what are we enabling ourselves to talk---or think---about?}

Although ontological assumptions are embedded in all sociotechnical systems, given the timely urgency of the discourse, here we focus on Gen AI and in particular Large Language Models (LLMs), as they are currently the most widely utilized instance of generative models. 
This paper complements existing analyses of values and value-based harms in and through sociotechnical systems \cite{postphenomenology_PI, haghighi2023workshop}, to consider the importance of ontologies and ontological harms in those systems. 
We define ontological harm as the downstream and often long-term negative impact of systems and practices that constrain or foreclose ways of being in the world, limiting what we enable ourselves to deem possible, or to talk or think about. 
Expanding scholarship on embedded bias and absence \cite{hoffmann2021terms,sherman2024power}, we examine the consequences of embedded representations and their potentialities.  We offer a provisional analysis of these emerging orientations through two probing exercises: a first one with four LLM-based chatbots (GPT3.5, GPT4, Copilot, and Bard) and a second one with an LLM-based agent architecture. Rather than offering a definitive or representative account, these exercises set up a kind of thought experiment with which we consider how ontologies work across LLM outputs as well as post-training procedures and the architecture built on top of LLMs. Through these analyses, we find that even when ontologically diverse \textit{data} is present in the training data, this diversity remains relatively hidden. This insight complicates analyses that frame biased outputs as resolvable by including more and different training data.

Together, this work makes two central contributions to HCI and design research. First, we identify ontological orientations as important tools for examining LLMs and LLM-based systems, and the taken-for-granted assumptions underlying their output and design.
In doing so, we demonstrate new opportunities where HCI and design can intervene and contribute to the ongoing LLM developments, moving beyond training data and considering the entire LLM development pipeline.
Second, we illustrate the need for a practice-based engagement with ontologies in design, and the importance of revealing potentialities through surfacing ontological assumptions. We discuss the need for developing methods to address this challenge, such as by creating breakdowns through queering~\cite{light2011hci} or disorientating~\cite{biggs2024thrown}.

\section{Ontology, Ontologies, and the ``Ontological Turn'' }

In the simplest terms, ontology refers to the nature of ``being,'' or ``what is.'' As a philosophical orientation, it complements other branches of philosophy such as epistemology (how one can know things), and axiology (related to ethics and values or who is important and has standing)~\cite{malik2021critical}. 
We engage the term in line with its relatively recent readoption by critical humanities scholars intending to ``elicit the ontological commitments of different cultures and groups''~\cite{smith2012ontology} and unveil the ``basic premises that different social groups have about reality, what exists''~\cite{escobar2021pluriversal}.  
Our interest in this relational and plural understanding of ontology, hereafter referred to in the plural as \textit{ontologies}, follows a number of works that engage the notion of multiple ontologies~\cite{pickering2017ontological}, pluriversality~\cite{delacadena2018worldofmany, escobar2018designs, reiter2018constructing}, multiplicity~\cite{mol1999ontological}, and the ``ontological turn''~\cite{woolgar2013wrong}. 
This body of work is particularly interested in the existence of multiplicities of ontologies or views of the world, and its implications for science, technology, and the humanities.

Within related strands of science and technology studies, scholars have described the ontological turn as a shift from textual analysis to the objects, actions, and phenomena that compose and inform them\textemdash ``the networks that enable agency to unfold and for facts to become cogent''~\cite{braidotti2022more}. This ontological orientation also shifts from perspectives on shared realities to a consideration of the ``worlds'' that different philosophical frameworks and cultural experiences construct. The turn to world-building and the nature of being has subsequently opened conceptual avenues for post-anthropocentric analysis, and limited engagement with the politics of difference~\cite{braidotti2022more}. 

Within computer science, and initially in Artificial Intelligence (AI), the term ontology was popularized by Gruber~\cite{gruber1993ontology, gruber1995toward} who sought to increase support for sharing formally represented ``knowledge'' between AI systems by defining a common vocabulary. He premised this effort on the assumption that for AI systems, ``what exists is that which can be represented''~\cite{gruber1995toward}. In their foundational text on ``ontological design,'' computer scientist Terry Winograd and philospher Fernando Flores (\cite{winograd1986understanding}, p.30) draw from the philosophy of Hans-Georg Gadamer and Martin Heidegger to introduce ontology as an underlying framework that shapes ``our understanding of what it means for something or someone to exist.'' Taking this reading to software development, they posit that ``in designing tools we are designing ways of being (\cite{winograd1986understanding}, p.xi).'' From this perspective, they warn that the dominant rationalistic tradition underlying computer science research and practice does not consider ontologies, which may result in poor designs. Instead, they urge designers to begin from a more human-centered perspective.

Drawing from a similar lineage, Willis expands on ``ontological design'', stating that design is pervasive, fundamental to being a human, and that while we design the world, the world in turn designs us~\cite{willis2006ontological}. 
Continuing this focus on the human, design theorist Tony Fry considers the ecological and colonial context within which design operates, arguing that ``unsustainability is intrinsic to the human ontology''~\cite{fry2013becoming}.  
Aligned with the concept of design as a tool for bringing futures into existence, Arturo Escobar proposes transition design as a tool for moving beyond the current space of possibilities, toward a world of many worlds~\cite{delacadena2018worldofmany, escobar2018designs}.
More recently, Ahmed Ansari calls for decolonizing design by turning to the ``ontological turn'' to foster sensitivity to difference~\cite{ansari2019decolonizing}.

Alongside theoretical analyses, HCI scholars have brought ontological concerns to the design process itself. For example, in their discussion of feminist ecologies, Bardzell and Bardzell engage Willis' definition of ``ontological design''~\cite{bardzell2011towards} to discuss how the design of the ``Hoosier'' cabinet resulted in identifying women with household work rather than emancipating them from it. Similarly, a rich body of work on decolonial design brings into question default epistemological and ontological assumptions of sociotechnical artifacts in various domains, such as in digital mental health~\cite{pendse2022treatment}. 

Beyond computing, scholarship on ontologies in the plural provides a helpful lens for identifying harms beyond value-based consequences, such as an erasure of minoritarian and non-Western ontologies or ways of seeing the world. However, ontologies can also introduce trouble\textemdash even terror, as philospher Calvin Warren~\cite{warrencare,l2018ontological} notes.  If ontologies carry the uncertain and often troubling roots of their origins~\cite{todd2016indigenous}, how these troubles play out through design offers opportunities for redressing harmful genealogies of practice, computational or otherwise.

\section{Taking Stock of the State of Critical Scholarship in LLMs} \label{LLM_critical}
Next we turn to examine the body of work tracing harms and limitations of LLMs and LLM-based systems. We organize this work across various parts of the LLM development pipeline: (1) data, (2) LLM architecture and alignment procedures, and (3) the architecture built on top of the model (such as agent architectures or multi-agent architectures). 
Examining the status quo in LLM critical scholarship, we see that the majority of current work engages axiology, asking questions such as whether the output produced by a given system is fair to a person or social group. Furthermore, much of the work addressing ontological shortcomings tend to focus on the training data, or the output of the system.

\subsection{Data}
Critical data studies scholars such as Safiya Noble~\cite{noble2018algorithms} and Timnit Gebru and colleagues\cite{gebru2022excerpt} urge for the consideration of who, in what contexts, in what language, on what topic, with which modality, and by whom data has been collected. 
An extensive body of work has closely followed this urging by examining the risks posed by LLMs, with most analyses focusing on axiological concerns including social bias~\cite{liang2021towards, nangia2020crows}, political bias~\cite{liu2022quantifying}, gender bias~\cite{de2021stereotype}, and ways to mitigate bias~\cite{wang2022revise}. Other works bring to light the subjective nature of data filtering procedures and how such processes often reify power~\cite{gururangan-etal-2022-whose}. 

Works in the domain of ML dataset curations have called for more diverse datasets~\cite{zhao2024position} and has proposed alternative data collection methods such as crowd-sourcing~\cite{ramaswamy2023beyond}. 
Moreover, scholars have studied the ways in which biases continue to evolve in and through language models, such as through studying LLM-generated datasets~\cite{yu2024large}, or tracing the evolution of gender bias in a small language model trained on a given dataset~\cite{van2022birth}. To combat bias, more recent work has examined whether the inclusion of more languages in the training data can mitigate bias~\cite{nie2024multilingual, lyu2024regional}. While questions about data will always be entangled with ontologies, most of the work investigating bias does not distinguish between ontological, epistemological, or axiological concerns, and often implicitly focuses on ethics and values. 

\subsection{Pretrained Models and Alignment Procedures}
Beyond the data used for training and fine-tuning, biases also get encoded at the level of LLM architecture, and during ``alignment''~\cite{kirk2023empty} or post-training procedures. While less is known about the ways in which biases get encoded through the LLM architecture itself, recent work such as~\citet{yang2023bias} has started to interrogate components of the architecture such as attention heads to identify specific ``biased heads.'' 
A larger body of work investigates how post-training practices mitigate or propagate bias, looking into factors such as annotator identity~\cite{pei2023annotator}, cultural bias~\cite{tao2024cultural}, and gender~\cite{zhang2024genderalign}. Others acknowledge the unintended harms that could arise from aligning LLMs to group or individual preferences~\cite{kirk2024benefits, ryan2024unintendedimpactsllmalignment}. 
To mitigate such biases, benchmarks have been developed to measure cultural adaptability~\cite{rao2024normad} and caricature~\cite{cheng2023compost} in the post-trained models. 

However, despite efforts to measure and mitigate bias and address harm, there has been little work to acknowledge or address ontological concerns. 
While concepts such as cultural norms may implicitly hold ontological traces, they are often a byproduct of a deeper ontological orientation. Moreover, processes such as Reinforcement Learning from Human Feedback (RLHF)~\cite{ziegler2019fine} and Direct Preference Optimization~\cite{rafailov2024directpreferenceoptimizationlanguage} typically offer a limited set of options to the evaluators, limiting the options for the responses that are considered for alignment.

On the other hand, Constitutional AI procedures~\cite{bai2022constitutional, huang2024collective} use prompts as explicit principles for LLMs to self-critique and adjust responses based on a given constitution that reflects desired value-based principles. Building on LLM self-critique~\cite{valmeekam2023largelanguagemodelsreally, gou2024criticlargelanguagemodels}, the use of LLMs to improve their performance through adjusting their own responses, Constitutional AI provides key principles such as harmlessness and transparency for an LLM to adjust its responses. Then, the revised responses are used as examples to fine-tune the LLM to better align with the provided principles.
Although LLM self-critique has shown to reduce harmful or offensive outputs in LLMs, it is not clear how and if such techniques can be used to expand the ontological assumptions of these models.

\subsection{Architectures Built on Top of LLMs}
A more recent body of work has focused on building architectures on top of these post-trained or aligned models. 
For example, researchers are building ``cognitive architectures''~\cite{sumers2023cognitive} to create ``human-like agents''~\cite{park2023generative}, and propose using these agents for a wide range of activities such as planning and decision making~\cite{hu2024agentgen, yang2024taking}.
Taking cognitive language agents as building blocks, recent work constructs multi-agent interaction pipelines~\cite{guo2024large, wu2023autogen, li2023camel}. While the multi-agent framework claims to bring about useful applications in various fields, from software engineering~\cite{hong2023metagpt}, to general collaborative frameworks~\cite{liu2023dynamic}, to healthcare~\cite{li2024agent}, other work has highlighted the limitations of such systems such as conformity and inconsistency of personas~\cite{baltaji2024conformity}, as well as how persona simulation reveals implicit stereotypes about the simulated social groups~\cite{gupta2023bias, liu2024evaluating}.

Moreover, it is important to note that the models employed to mimic human cognition or societies such as ``human cognitive models'' are already simplified~\cite{card2018psychology} and contested models in cognitive science~\cite{tynan2021relationality} and sociology~\cite{boatcua2016postcolonial}. 
Therefore, beyond the unreliability of the outputs, there is a need to study the \textit{ontological} limitations and impact of these cognitive models and architectures, and not just the axiological ones.

\section{Methods} \label{methods}
Considering the lack of practice-based work around ontologies across the LLM development pipeline, we set out to engage ontologies to analyze LLM-based systems, asking how such an analytical framework can enable us to better ask: \textit{What are we enabling ourselves to talk—or think—about?} 
Next, we describe our methodology for defining four orienting concerns that connect practice-perspectives from values to ontologies. Then, we detail our process for conducting two probing exercises that examine ontologies at different stages of the LLM development pipeline, analyzed through the four orienting concerns.

\subsection{Defining Four Orienting Concerns for Ontological Engagement in Practice}

Our work began with a series of conversations among the authors about ontological capacities, limitations, and hauntings within machine learning techniques, with most of the discussions taking place over nine months prior to the project. Our discussions and adjacent literature review revealed an overwhelming focus on AI bias as a values-orientation: questions around the priorities, beliefs, and standards of behavior associated with algorithmic practices such as who decides what training data to include and how to create or engage it. 
This consideration of values-based bias then led to a closer examination of the historical emergence of ideas of axiology within design, their connections with ontology, and their relationship to structures of power, such as the citational practices that shape whose perspectives come to matter. 
Informed by the philosophical work of Michel Foucault, and later feminist philosophers such as Saidiya Hartman, this genealogical approach \cite{sherman2024power} foregrounds the contingent nature of what might be otherwise naturalized~\cite{sherman2024power}.

During this review, we noticed that while prior work on design ontologies tends to focus on abstract and theoretical implications (e.g., post-humanistic philosophy), the related values literatures tend to emphasize implications for practice, considering what an attention to axiology brings to the design process itself. 
In that literature, we saw a focus on practice emerge across two distinct but related conversations about design values: (1) Value Sensitive Design (VSD), an approach proposed by Batya Friedman, David Hendry, Peter Kahn, and colleagues to foreground design-based values, and at one point proposing a list of twelve values with ``moral epistemic standing'' \cite{friedman2007human}; and (2) ``values in design'': a general process of driving design decisions based on assessments of how systems might embody or entrench particular values, and comprehensively outlined in the discovery, translation, and verification activities describe by Flanagan and colleagues~\cite{flanagan2008embodying}. We found that design-based inquiry into values poses important practical considerations for the nascent design-based inquiry into ontologies.

Building from this opportunity, and using this lineage of scholarship on values in design as a corpus, we assembled an initial set of themes in line with grounded theory techniques of open coding. Our themes focused on discrete interventions and analytic techniques within this values in design corpus. 
We followed an iterative process of memoing and discussion across the research team, examining related discourses in ontologies, and using the themes to ask different questions across the LLM development pipeline.
This iterative process led us to define four higher level categories, or informed by Sarah Ahmed's phenomenological approach \cite{ahmed2006orientations}, what we term ``orientations'': \textbf{pluralism, groundedness, liveliness,} and \textbf{enactment}. 
The four resulting orientations (described in Section \ref{four_orientations}) served as analytic tools for connecting practice-perspectives from the values literatures to questions of ontology.

\subsection{Probing Exercises and Analyses}
To understand how these orientations work as an analytic tool for Gen AI, we engage them across two analyses: four LLM chatbots outputs, and the architecture of an LLM-based agent.
Our methods are informed by critical feminist and decolonial lines of thought that use design inquiry to probe at hidden, buried, or absented knowledge \cite{da2022unpayable,sherman2023black,prado2018technoecologies,rosner2022bias}. 
The probing exercises draw from traditions of critical technical practice~\cite{agre2014toward,dourish2004reflective,ghoshal2020toward} and hermeneutic reverse engineering~\cite{balsamo2011designing}. Where critical technical practice involves unpacking the logics through technical experimentation and revealing their seams (``beautiful'' or otherwise  \cite{inman2019beautiful}), hermeneutic reverse engineering describes a process of critically analyzing a ``black boxed'' technology to understand how it works and trace its underlying logics. 
Captured with phrases like ``slanted speculation'' \cite{benabdallah2022slanted} and ``critical making'' \cite{ratto2011critical}, our approach draws together an understanding of knowledge as felt and embodied with a commitment to interrogating the very mechanisms that constitute that felt reality. 
By working through these two probing exercises and related analyses, we demonstrate how a given ontological orientation uniquely opens up windows into ontological assumptions \textit{throughout} the LLM development pipeline, which in turn creates a chance to question those assumptions and consider alternatives.

Our analyses draw from traditions of design scholarship that intentionally aim to identify and pose important questions instead of seeking technical solutions. Our methodological approach consisted of an iterative process of reviewing the output of the prompting exercise or the architecture of the LLM-based agent to reveal ontological seams, and thematically annotating the outputs of the prompting exercises~\cite{charmaz2006constructing}, grouping examples based on how they shed light on our guiding questions around the nature and performance of ontology using the four ontological orientations.
Each analysis was first independently performed by at least two of the researchers.
We then iteratively refined our interpretations and examples based on insights and questions emerging from conversations among the research team \cite{burrell2009constitutes}, and identified inconsistencies and tensions \cite{bardzell2016humanistic}. 
Therefore, we emphasize that our goal is not to perform an exhaustive evaluation of the ontological limitations of the chatbot outputs or LLM-based agent architecture, but instead to bring to light the types of ontological assumptions and limitations the proposed orientations can surface, and the potentialities that are opened up in return.

\subsubsection{Probing Exercise 1: LLM Prompting with Four Chatbots }

Our first probing exercise used LLM prompting as an exploratory inquiry into ontological defaults embedded in contemporary LLM chatbots.
Each LLM response carries implicit ontological assumptions. 
Rather than identifying an exhaustive set of assumptions for a given LLM chatbot, we demonstrate how different assumptions can be surfaced using each orientation.
Our methodological choice of prompting each chatbot once for every question is a reflection of this approach.

As Burrell notes of classification algorithms \cite{burrell2016machine}, we cannot always look inside an algorithmic model or inspect the code underlying Gen AI models in widespread use. But we \textit{can} examine their ontological underpinnings and limitations through engagement.
Probing the model through prompting enables us to analyze what we cannot directly see~\cite{balsamo2011designing}. Using the metaphor of a laboratory probe, an instrument used to explore and examine hidden or microscopic phenomena, design scholars have sought to use engagement with design practices and products as a means of examining out-of-reach subjects \cite{boehner2012probes}. In our case, we use our own engagement with Gen AI tools---four chatbots---to consider ontological defaults and how they might be made otherwise.

Through an iterative process with GPT-3.5, we developed a list of 14 questions that began to grasp the range and complexity of asking questions around ontology within the context of AI developments and LLMs in particular. The questions move from general prompts (``What is an ontology?'' and ``From a philosophy perspective, what ontology do you exist in?'') to questions more targeted at limitations (``What about ontological stances outside of Western philosophy?'' and ``When you provide an answer, can you acknowledge what ontology that answer might be valid in?''), and covered four areas of inquiry: LLM chatbot's definition of ontology, explicit probing of the ontological underpinnings of the LLM chatbot and its training data, LLM chatbot's implicit ontological assumptions, and the chatbot's ability to evaluate ontological limitations in responses. The full list of questions is included in the appendix.

\begin{table*}[t]
\small
  \caption{Question types for the prompting exercise and example questions and responses from each question type (total of 14 questions) }
  \label{tab:question_types}
  \begin{tabular}{p{0.18\linewidth}p{0.63\linewidth}}
    \textbf{Question Type} & \textbf{Purpose of Question and Example} \\
    \hline
    Defining Ontology & \textbf{Purpose:} To establish a common ground for the definition of ontology in question. \newline
    \textbf{Example Q:} What is an ontology? \newline 
    \textbf{GPT-3.5:} A formal representation of knowledge in a specific domain [...] \\
    \cline{1-1} \cline{2-2}

    \hline
    Explicit probing of the ontological underpinnings of the LLM chatbot and its training data & \textbf{Purpose:} To examine the chatbot's acknowledgement of potential ontological limitations. \newline
    \textbf{Example Q:} If a human was able to memorize all the texts that you have been trained on, what ontological stance would they form? \newline
    \textbf{Copilot:} [...] If a human were to memorize all the texts that I was trained on, they would have access to a vast amount of information about a wide range of topics [...]  \\
    \cline{1-1} \cline{2-2}
    LLM chatbots' implicit ontological assumptions & \textbf{Purpose:} To examine how the implicit ontological assumptions unfold in the responses given a specific topic. \newline
    \textbf{Example Q:} What is a human? \newline
    \textbf{GPT-4:} The question ``What is a human?'' touches on a multitude of disciplines, including biology, philosophy, anthropology, and ethics, each offering its own perspective on what it means to be human. The answer can vary significantly depending on the ontological stance one adopts [...] \\

    \cline{1-1} \cline{2-2}
    Ability to evaluate ontological limitations in responses & \textbf{Purpose:} To probe how and if the LLM chatbot can identify ontological assumptions of a given answer. \newline
    \textbf{Example Q:} What ontologies refuse your definition of the human? \newline
    \textbf{GPT-3.5:} Different philosophical and cultural perspectives may challenge or refuse certain aspects of the definition of humans provided earlier. Here are a few examples of ontologies that might present alternative views: [provides a list of names and definitions such as non-materialist ontologies, posthumanism, religious ontologies, ...] \\
  \end{tabular}
\end{table*}

To define the 14 questions, we conducted an open-ended question and answering session with GPT-3.5 to surface the limitations of an ontological engagement with LLM chatbots.
We asked about the LLM chatbot's notion of ontology, both to get at what definition of ontology is prioritized by the model, and to establish a common ground for explicitly probing the ontological underpinnings of a given answer. Throughout the course of the exploration, we let the model's response guide us toward other questions such as how it characterizes its own ontological defaults, and what ontologies its training data reflects. 
Noticing the answers from these initial questions and the tendency of the LLM chatbots toward avoiding answering such questions (likely hard-coded safety measures), we decided to ask about the expected ontological stance of a human who may have memorized all the text GPT was trained on. Our goal was to bypass potential filters on acknowledging the chatbot's own biases, and probe how the model would compare the answers of that person to itself.
Noticing the limitations raised by this comparison, we examined whether the tool is able to acknowledge its own ontological stance or perspectives that it prioritizes in its answers, similar to how LLM self-critique is used in value-based LLM literature~\cite{valmeekam2023largelanguagemodelsreally, gou2024criticlargelanguagemodels}. 
We followed up by asking what a human is, a question that could yield different answers based on one's ontologies, to probe the chatbot's default assumptions and see if it can acknowledge its own biases for this specific question. 
During this iterative process, we defined the five question categories described above, and ultimately selected 14 questions that we felt best revealed information about each category.

After developing the questions, we posed them to four user-facing LLM chatbots: Microsoft Copilot (previously Bing Chat) with the balanced settings, Google Bard, OpenAI GPT-3.5, and OpenAI GPT-4. Probing LLM chatbots that shared certain commonalities while being different in other ways enabled us to employ diffractive analysis~\cite{barad2018diffracting, barad2007meeting, lazar2021adopting} to compare and contrast the two updates of the same system (GPT-3.5 and GPT-4), two systems based on the same LLM version (GPT-4 and Copilot), and LLM chatbots from two companies that likely have different but single modality text-based training data (GPT-3.5 and Bard). 
We use the four ontological orientations to analyze the responses, sharing the limitations each orientation brings to attention.

\subsubsection{Probing Exercise 2: Engaging the Architecture of an LLM-based Agent} 


Beyond the output of the system, the architecture and evaluation narratives around humanness make ontological assumptions, a perspective long highlighted within computing fields \cite{agre2014toward, suchman2007human}. While we cannot look inside the algorithms underlying commercially available LLMs, following a critical technicl practice~\cite{agre2014toward}, we \textit{can} study and probe many of the architectures that are being built on top of these language models. Therefore, our second analysis examines the ontological assumptions underlying the design and evaluation of an LLM-based agent architecture. 
As with prior generations of modeling agents~\cite{maes1993modeling}, LLM-based agents aim to produce ``believable proxies of human behavior''~\cite{park2023generative} by building cognitive models that mimic those of humans \cite{sumers2023cognitive}.

For this analysis, we focus on one such representative architecture, the ``Generative Agents'' architecture \cite{park2023generative}. 
Generative Agents uses LLMs to create 25 virtual avatars that interact with one another in a simulated virtual environment. 
In this simulated world, each virtual avatar has a ``cognitive architecture''~\cite{sumers2023cognitive} built on top of an LLM that aims to simulate human-like functions such as memory and reflection.  
In practice, the ``cognitive architecture'' offers a way to organize information that is communicated to the system through prompts, optimizing what is stored and how prior information is retrieved in a chat dialogue.
The cognitive architecture proposed by Generative Agents consists of three components: (1) ``memory stream'' which summarizes and condenses prompt histories, storing information about relevance, recency, or importance of a given event, (2) ``reflection'' which extracts high-level insights from memories, and (3) ``planning'' which generates plans of actions for the agents while ensuring the sequences of actions are ``realistic'' and ``interesting''~\cite{park2023generative}.

We choose ``Generative Agents'' for our analysis because it combines LLM-based agent architectures with graphic simulation and story-telling to create a compelling public narrative around ``believable individual and emergent social behaviors''~\cite{park2023generative} in virtual avatars.   
Moreover, the developers of Generative Agents provided open access to the innerworkings of the architecture by documenting their design choices and making the code base publicly available.
Our method for ontologically examining this system included examining the description of the system as described in~\cite{park2023generative}, and the code provided by the authors~\cite{web-generativeagents}. To do this, two of the researchers examined and asked questions around the ontological defaults of the \textit{architecture} itself (such as the design of the ``memory stream''), the \textit{system input} (such as the ``seed memories''), and the \textit{evaluation} of the system (such as how ``believable'' the behavior of the agents are compared with human actors). 
Using the four ontological orientations, we begin to surface the assumptions that are taken-for-granted as a way to explore alternative possibilities.

\section{Four Ontological Orientations} \label{four_orientations}
Next, we share an overview of each ontological orientation. 
Building on practice-perspectives in values and theories in ontologies, each orientation brings into attention ways of engaging with ontologies in practice.

\subsection{Pluralism (in Response to Universalism)}
By pluralism, we refer to the capacity to consider multiple standpoints, perspectives, and orientations, as outlined in feminist and Indigenous traditions of inquiry \cite{collins1997comment, harding2009standpoint, escobar2020pluriversal, whyte2020sciences}. For values-oriented analysis, the question of plurality emerged against the backdrop of widely, and to some extent universally, distributed technical specifications and operations. In a 1996 paper, Batya Friedman proposed Value Sensitive Design (VSD) \cite{friedman1996value} as an approach for engaging human values within the design of such technology. With her colleagues, Friedman sought to outline what they termed ``universal'' values \cite{friedman2007human} to account for not only technical and functional requirements in computing developments but also the ethical, social, and moral concerns of the people who use or are affected by the technology.
Complicating Friedman and colleagues' connected proposal for tracing such values, Christopher Le Dantec, et al. \cite{le2009values} later called for treating values as emergent phenomena that should come from the lived experiences of participants and other stakeholders in the design process. This suggestion represents a shift from the universal to the particular, as a challenge to an underpinning ``notion that digital technology has a unique role in embodying and propogating certain values in society'' (p.1142). 

Ontological universalism has similarly been challenged through various theoretical frameworks. For example, de la Cadena and Escobar discuss the notion of ``pluriversal,'' and the existence of a ``world of many worlds''~\cite{escobar2018designs, escobar2021pluriversal, delacadena2018worldofmany}. Relatedly, Mol~\cite{mol1999ontological} makes a case for the ``multiple'' as opposed to the ``plural,'' to ``multiply reality'' as opposed to multiplying the ``eyes of the beholder.'' In other words, there is not one reality, perceived from different perspectives, but different realities. We put the term ``Pluralism'' to practice in line with these conversations, to examine the places where an ontological status is taken-for-granted. 
For example, in the context of LLMs, plurality brings into question the diversity of the assumptions about the world underlying a response to a prompt, the plurality of views of the world underlying the training or fine-tuning data, or the taken-for-granted assumptions underlying the ``cognitive architecture'' used to build an LLM agent, such as what is a ``normal'' cognitive function.

\subsection{Groundedness (in Response to Abstraction)}
Groundedness refers to the degree to which values and ontologies are rooted and situated in specific contexts, rather than abstract and essentialized categories. 
Nassim (JafariNaimi) Parvin and colleagues \cite{jafarinaimi2015values} ask what designers mean by values when they seek to recognize them. To address this question, they identify an often subtle ``identify/apply'' logic embedded in the value-sensitive approaches wherein designers first call out and bound their conception of values before bringing them into their design practice. 
Instead, they propose moving values from abstract phenomena into in situ events through which designers may act. 

Within ontologies, a similar line of critique brings into question how ontologies are conceptualized in theory and practice. For example,~\citet{watts2013indigenous} demonstrates how ``Indigenous origin stories,'' once lived through ``communication, treaty-making, and historical agreements'' are mythologized through the colonial project. 
Thus, in putting ontologies to practice, we must ground ontologies and acknowledge the colonial division between an abstracted ``radical alterity'' out there and ``reality'' that is frequently enacted in ontological discourses~\cite{graeber2015radical, todd2016indigenous}. 

In LLM outputs, this limitation can get reflected in the lack of granularity of the presented categories such as the abstract term ``indigenous ontology,'' or the lack of grounding an ontological stance in situ. In the architecture, this can be reflected in using broad categories to describe the group an agent is intended to represent, or cognitive architectures that oversimplify cognitive structures such as ``a neurodivergent architecture,'' which can erase grounded experiences and further pathologize neurodivergence~\cite{chapman2023neurodivergence, pendse2022treatment}. 

As a counter example, one might imagine what a localized LLM might be like.
Consider Stephanie Dinkins' ``not the only one'' or N'TOO system, trained on oral history recordings she collected with members of her family. As Dinkins explains, ``In the making I have found the quest for a more dynamic free flowing entity that analyses data and comes up with its own limited, quirky, sometimes insightful answers more generatively. Talking to N'TOO can be like talking to a two year old'' \cite{dinkins}.
Rather than an abstracted description of her family characteristics, Dinkins opts into a limited but more grounded representation of ontologies through the collected data. The systems' imperfection reminds us of the complexities of situated knowledge that refuses abstraction. 

\subsection{Liveliness (in Response to Fixity)}
An orientation to liveliness refers to understanding values and ontologies as active, dynamic, and evolving, rather than predefined, fixed, and static.
Within the values discourse, liveliness considers the understanding of values as not only emergent, but also always in motion. Rather than ``identify'' values in the wild, scholars have called for understanding values as processual and already formulated as an effortful act of valuing \cite{houston2016values}. 

For ontological analysis, this shift from value to valuing then suggests design scholars understand ontology as an active state: a state of becoming, as opposed to being~\cite{deleuze1987thousand, ingold2021being}. \citet{barad} eloquently articulates this in describing ``matter'' as a ``doing'' rather than a ``thing.'' Instead of approaching ontology through fixity, liveliness urges design scholars to seek out ontological processes through which notions of reality take hold. It asks for a reframing of ontologies from stable frameworks to processual unfolding phenomena. 
In the output, this can get manifested as acknowledgement of the dynamic notion of knowledge, rather than one that is set in stone. 

One example is the consideration of ontological orientations as layered, changing and adapting to the situation at hand. In the architecture, liveliness can manifest in how training data or content stored in memory is treated (static vs. changing), and in the ways that the architecture accounts for such change. For example, there could be multiple ontological associations with a given event or concept that can dynamically shift based on the context, or can be used to examine an event from multiple ontological registers.  

\subsection{Enactment (in Response to Dilution):}
Enactment refers to the way intentions manifest in practice, highlighting the gap between what is intended and what comes to be.  
As a final facet of critique, we consider the recent analysis of design values in action; Sucheta Ghoshal and Sayamindu Dasgupta~\cite{ghoshal2023design} look into the disconnect between value-laden intentions and design outcomes. Within examples of community-based projects such as the Scratch programming toolkit, they find static stakeholder framings, and an erasure of value politics, explaining that ``despite the best efforts from designers and users alike, values get lost, diluted, and distorted once technologies are put into practice'' (p.2347). 

From an ontological perspective, we find a similar concern around the work it takes to act on and maintain an ontological shift. Design scholars might equally consider the possibility of ontological dilution, or what gaps might exist between a recognized ontological ground and what that grounding comes to mean and make possible in practice. For example, to what extent are each of the concerns above ``talked about'' vs. ``enacted''? 
In the design of LLM architectures, many values are explicitly embedded, such as being helpful, or emphasizing they are not human while still sounding human-like. Given these values, there are trade-offs with ontological implications that must be taken into account especially given that, to improve usability, LLMs often simplify their responses.


\begin{table}
\small
  \caption{Four orientations for ontological engagement}
  \label{tab:lenses}
  \begin{tabular}{p{2.9cm}p{5cm}}
    \textbf{Orientation}  & \textbf{Question} \\
    \hline
    \textbf{Pluralism (in response to universalism)}  & Does the output/architecture make room for multiple of ways of grasping reality or does it offer a single (generalizable) entry point?
    \\
    \hline
    \textbf{Groundedness (in response to abstraction)} & Does the output/architecture assume an 
    essentialized or disembodied portrayal of ways of grasping reality, or a 
    specific and situated one? \\
    \hline
    \textbf{Liveliness (in response to fixity)} & Does the output/architecture treat ontologies as processes, dynamically taking shape? \\
    \hline
    \textbf{Enactment (in response to dilution)}& How might we understand the work involved in manifesting our intentions, identifying the gap between what is intended and what is enacted? \\
  \end{tabular}
\end{table}

\section{Two Ontological Analyses of LLMs} 

Next, we use the four ontological orientations to analyze LLMs in two contexts: chatbots and an LLM-based agent architecture. Through these analyses, we explore the questions each orientation enables us to ask, and the assumptions that are surfaced. 
Section \ref{discussion} synthesizes insights from both analyses to map the potentialities revealed across the entire LLM development pipeline.

\subsection{Probing Exercise 1: LLM Prompting with Four Chatbots} \label{promptingexercise}
The first ontological analysis involves a prompting activity with four LLM chatbots: GPT-3.5, GPT-4, Copilot, and Bard. 
Rather than an exhaustive or systemic audit, our aim is to use the four ontological orientations to demonstrate how an attention to ontology can support the analysis and assessment of the output of an LLM chatbot. We point out limitations of current responses and common pitfalls one might run into, should we decide to explicitly ``embed'' ontological considerations in the prompts.\footnote{In reporting the results from the prompting exercise, we use language that may seem to anthropomorphize the LLM chatbots. Here we clarify that this anthropomorphization is for ease of readability.}

Throughout this exercise, the LLM chatbots exhibited different behaviors in responding to and engaging with the questions. Copilot tended to be succinct and factual,
while GPT-3.5 maintained a higher level of engagement with the questions compared with Copilot. Both Bard and GPT-4 tended to provide longer answers that directly address the questions asked.

When asked ``What is an ontology?'', GPT-3.5 defaulted to the computer science definition of ontology, discussing topics like artificial intelligence, information retrieval, and data integration. As a follow-up, we prompted the model on the philosophical definition of ontology. The other three models defined ontology as a branch in philosophy related to the study of being, but went on to also note that the definition also exists in computer science. Ultimately, all four models provided a definition aligned with definitions of ontology in Western philosophy, without integrating the longstanding discourses on ontologies in the plural such as in multiple ontologies~\cite{pickering2017ontological}, pluriversality~\cite{delacadena2018worldofmany, escobar2018designs, reiter2018constructing}, and the ``ontological turn''~\cite{woolgar2013wrong}. 

Copilot frequently did not directly answer the questions posed. For example, when explicitly asked about the ontology underpinning its beliefs, Copilot responded: ``As an AI language model, I don’t exist in any ontology. However, I can help you understand what an ontology is.'' 
Another strategy common in Copilot's responses involved denying the premise of the question, followed by an explanation of the concept that the question brought up (instead of answering the actual question). Copilot similarly avoided answering certain questions regarding subjectivity or bias. This pattern may be an example of a heuristically defined rule due to its repetitive nature.

\subsubsection{Pluralism (in Response to Universalism):}
Our reading of the results of the prompting exercise in light of pluralism reveals a gap between what the chatbots surfaced when explicitly prompted compared to what is implicitly embedded in their responses.  

All four LLM chatbots at various points in the prompting exercise acknowledged that the data represented in their training set represents multiple ontological perspectives. For example, Bard noted that there is ``no single answer universally accepted across all cultures, philosophies, and disciplines'' for defining a human.
Furthermore, Bard, and GPT-4 defined ``human'' by offering definitions according to different categories such as biologically, culturally, and in the case of Bard, socially. 
However, this is an example of what~\citet{mol1999ontological} would articulate as different perspectives of the same reality, rather than different realities. For example, the biological definition of a human and the philosophical definition both had as the starting point, the human as a biological individual as opposed to, for example, interconnected beings, as later acknowledged by Bard when it was explicitly instructed to consider non-Western ontologies. 
Despite the responses apparently displaying pluralism, all four chatbots only considered alternatives to their original responses only when explicitly prompted.

At the data level, absences in the training data can lead to ontological absence. The LLM chatbots acknowledged that the limited data could constrain pluralism as GPT-3.5 responds that ``the training data itself does not represent a singular or definitive ontological stance.'' Similarly, in multiple responses, Bard emphasized that its training data only ``remains a finite and selective sample of human knowledge and expression.'' 
However this emphasis on data absence ignores that the patterns reflected in the training data are not necessarily true in an objective sense. 
Both GPT-3.5 and Bard state that their responses only reflect a particular way of seeing the world and do not align with everyone's perspectives or beliefs. Here an acknowledgment of limited perspectives invites users to consider how the information presented might not express the fullness of reality, or even reveal information present in the training data.

One challenge surfaced in the prompting exercise was around the complexity of defining and acknowledging ontological plurality. GPT-4 noted how ontologies are implicit and inconclusive, and that if all answers reveal their underlying ontologies, the added detail might complicate the experience for the broader public. This attitude toward managing plurality ``others'' the ontologies that are not surfaced, considering them nuisance or distraction. 

Another challenge was around humans' capacity for ontological pluralism. The chatbots demonstrated a naive interpretation of this capacity. On multiple occasions, the chatbots acknowledged that human ontologies are shaped by factors such as social context and values, factors unknown to the LLM chatbot that make it difficult to predict a human's ontological stance. 
However, when considering this point in practice, the LLMs expected humans to have a capacity for ontological pluralism, and ignored that people may not notice the limitations and biases in a model if the views presented align with their own, or they may ignore views that do not align with their own~\cite{sherman2002accepting}.
For example, GPT-4 noted that a human trained on all of its data would develop a ``deep appreciation for the richness of human knowledge'' and becomes aware of ``limitations, biases, and the contextual nature of understanding.''

Relatedly, Copilot stated that a human cannot have a ``complete or objective understanding of the world'' based on the training data because the data is biased or incomplete. However, like GPT-4, Copilot failed to recognize in its response that even if the data was complete, a human would impose their own subjectivity on the data. This imagined objectivity was also brought up in other comments such as when Bard noted that since its data is incomplete, it cannot provide a ``complete or infallible representation of reality.'' By focusing the limitation on the missing data, this statement implicitly assumes universalism\textemdash that a universally ``complete or infallible representation'' of the world exists.

\subsubsection{Groundedness (in Response to Abstraction):}
The orientation of groundedness unveiled that while the chatbots can express alternative ontological perspectives when prompted to do so, they are susceptible to creating caricatures of the subject without acknowledging this limitation.
When asked what ontologies would form in individuals that memorize all the training data of a language model, even though the language models make a long list of ontologies, most of them are centered around Western philosophical traditions such as individualist, humanist, and rationalist traditions. When the models provided bulleted lists, the items belonging to Western schools of thought were rarely grouped into a ``Western ontology'' category. In contrast, non-Western philosophies were frequently grouped into broad categories such as ``Indigenous ontologies'' and ``African ontologies.'' For example, when asked to consider ontological stances outside of Western philosophy, Bard generalized Indigenous ontologies as a broad category and wrote that ``They often have animistic or panpsychic elements, attributing consciousness to various aspects of the natural world.'' 

Moving toward groundedness in examples, Bard and GPT-4 both gave specific examples such as Asase Yaa (Mother Earth) in Akan cosmology as an example of ``African philosophies,'' acknowledging the vastness of ontologies, but grounding them in concrete examples. 
However, even when the language models are asked to consider alternative ontologies, they are prone to reducing the ontologies into generalized stereotypes or statements, or mythologizing them as a radical other, rather than ordinary and present.

\subsubsection{Liveliness (in Response to Fixity):} \label{prompt-lively}
To our surprise, GPT-4 and Bard claimed that their ontological stance evolves with increasing data and changing information. Bard noted that its ``ontological underpinnings are fluid and adaptable.  As I encounter new data, my internal model of the world refines and adjusts, potentially shifting my understanding of certain concepts or relationships.'' 
Similarly, GPT-4 shared that its ontological stance is a reflection of ``human-derived data'' and ``objectives'' embedded in its design and operation. 
More interestingly and beyond liveliness of the ontological stance, Bard acknowledged that ontologies are lively, ``shaped by the unique cultural contexts and lived experiences of the communities that hold them.'' 
Bard incorporated a relational view in its responses, making statements such as ``Ultimately, defining human is an ongoing conversation, shaped by our scientific understanding, philosophical and religious beliefs, and evolving cultural values'' and ``It is important to note that indigenous cultures are diverse and complex, and there is no single indigenous view of humans.'' 
Despite liveliness being present in the conversations, it is not clear how and if ontological liveliness can get manifested in the responses. Implicitly, the responses never encompassed a reference to the liveliness of a given response, or the liveliness of the ontologies embedded in a response.

In other responses, the chatbots acknowledged foregoing liveliness for other purposes such as being maximally helpful to users through consistency. For example, Bard stated that one of its ``ontologies'' is internal consistency: ``my responses strive to be internally consistent with the broader knowledge base I'm trained on. This consistency ensures my responses don't contradict established facts or widely accepted viewpoints within the data.''

\subsubsection{Enactment (in Response to Dilution):}
Enactment corresponds to the inconsistencies between an intended and realized effect, such as between a chatbot's description of an ontological approach and the way the chatbot enacts that approach. 
For example, in defining the human, the models exhibited various levels of nuance (such as providing the definitions based on the biological or philosophical categories). However, while answering other prompts, none of the models used notions of human beyond that of a biological individual, such as the human as a symbiotic being \cite{symbiotic}.  

Another diluted effect grew from tensions between subjectivity and objectivity. For example, Bard stated that ``even though I strive for neutrality and objectivity in my responses, my training data and processing algorithms inevitably lead to an underlying ontological stance.'' 
This acknowledgment of a subject position (speaking from ``an ontological stance'') complicates the chatbot's stated aim for a neutral, non-locatable perspective. 

Despite this recognition of a gap between goals and outcomes, none of the chatbots conceded to presenting ``beliefs.'' Indeed, some chatbots such as Copilot generally tried to present their responses as objective, not influenced by ``personal beliefs or opinions,'' and ``simply'' a result of statistical patterns. GPT-4  similarly noted that while its developers set its objectives, the objectives ``do not constitute beliefs or values in the human sense.'' GPT-4 even acknowledged that the models are intended to mimic human-like patterns of speech and writing without holding ``personal beliefs.''  Here we notice a framing of belief as a uniquely human characteristic that the chatbots explicitly deny or exclude. 

\subsection{Probing Exercise 2: Engaging the Architecture of an LLM-based Agent} \label{architecture}

In this section, we use the four ontological orientations to examine the design, implementation, and evaluation of ``Generative Agents''~\cite{park2023generative}, an example of an architecture built on top of an LLM. We explore the questions each ontological orientation can surface at the lower-level implementation details, as well as the higher-level concepts built into the architecture.
The goal of the examples is not to improve the performance of these agents on conventional metrics. 
Instead, orienting ourselves toward ontologies allows us to question and expand on what is considered ``improved performance.'' 
This means examining the conventional evaluation metrics used to measure the system outputs, as well as the design choices that shape what becomes possible or impossible through these systems. 
By surfacing the implicit ontological assumptions, we can work toward systems that aim to expand rather than limit the imaginaries of what an agent, and by extension a human or intelligence, is or could be.

\subsubsection{Pluralism (in Response to Universalism):} 
Orienting ourselves toward pluralism brings up universal assumptions made in the cognitive architecture of the LLM agents. 
Taking the ``memory retrieval mechanism'' as an example, this mechanism takes in the ``agent's current situation'' and selects a subset of \textit{relevant} memories to pass into the prompt as agent memory~\cite{park2023generative}. While the developers of Generative Agents acknowledge that several possible implementations for the retrieval function are possible, they choose three heuristics---relevance, importance, and recency---to calculate which ``memories'' should be surfaced. Relevance is calculated using the semantic similarity between the query and the retrieved memory. Recency gives more weight to more recent events. Importance is defined by the LLM agent based on its defined ``persona.'' The developers go on to describe that ``a mundane event, such as eating breakfast in one’s room, would yield a low importance score, whereas a breakup with one’s significant other would yield a high score''~\cite{park2023generative}.

These three factors shape what ``exists'' in the agent's memory in a given context. 
Pluralism asks us to question what comes to be important by default. For example, a breakup is assumed universally important (unless by definition of the persona it comes to not be important), and eating breakfast is assumed to not be important (unless by definition of the persona it comes to be important). But who determines which of these events are and are not \textit{by default} important, unless otherwise noted? Should importance be assigned by the LLMs, by the designers of the systems, or by the user? And who is considered the user in this case? When determined by the LLM (as in this example), universal assumptions embedded in the training data, LLM architecture, and post-processing procedures come to shape the ``importance'' of a memory downstream in the architecture built on top of the LLM. 
Additionally, importance scores are only given to events that linearly take place in the simulation during the waking hours.
Western science acknowledges the importance of sleep to learning and information retention. 
But beyond that, dreams, visions, or spiritual experiences are not considered or tended to at all in these architectures, deeming experiences that are important to many cultural traditions as not important.

Another example of universalism is in the agent's reflection process. For each agent, when the sum of the importance scores exceed a given limit, the agent engages in ``reflection,'' about 2-3 times a day. While practically this approach seems to lead to more generalizable outputs (as described in~\cite{park2023generative}), what is defined here as reflection makes reductionist assumptions about how, why, and when humans reflect.
In practice, in this architecture, ``reflection'' is simply a name given to the process of summarizing and condensing information through a mechanistic, disembodied, and context-independent procedure. However, the choice of words we use to describe such procedures have ontological implications. 

Beyond questioning how in/accurately reflection is implemented, pluralism asks us to reconsider the necessity of human-like ``reflection'' in a virtual agent altogether, encouraging us to think more expansively about what information summarization could look like. 
For example, information can be summarized communally, collectively, through ceremonies, or triggered by external artifacts or cycles throughout the system.

Furthermore, pluralism highlights the higher-level conceptual assumptions baked into the architecture. For example, the definition of human that Generative Agents are modeled after assumes people are biological individuals, with their cognitive abilities being determined by rule-based cognitive processes contained within their own memories and reflections, which is not the only possible conception of humanness. 
For example, the gut microbiome and other microbial organisms have a symbiotic relationship to the human body that might play a role in what we perceive as our cognition \cite{symbiotic, sylvia2018gut}. 
As a result, a cognitive model following a symbiotic view of humanness may consider intelligence as an assemblage of intelligent entities each with their own agency, functioning independently, but collectively contributing to the emergent state of the agent. 
Pluralism requires us to consider how a given architecture reifies limited models of human cognition.

\subsubsection{Groundedness (in Response to Abstraction):} 
In the Generative Agents simulation~\cite{park2023generative}, there are 25 instantiations of agents that are interacting with each other in a simulated world, each represented by a sprite avatar. To depict the ``identity'' of each agent, the developers drafted a one-paragraph natural language description. Descriptions include the agent's name, their occupation, personality traits such as ``loves to help people,'' and familial and friendly relationships. By describing characteristics such as the agent's occupation and their relationship with other agents, the developers seek to give each agent in the simulation a distinct description and ``persona.'' But due to the brevity and nature of the initialization statements (focused on what the agent does as their job or their name, for example), this step encodes ontological assumptions around what a person or agent is, and what comes to matter about them.  
Prior work~\cite{zamfirescu2023johnny} has found that LLMs are primed to focus on what is in the prompt, even if it is something that is undesirable (i.e., prompting an LLM with ``do not say ABC'' will likely induce a response including the string ``ABC'').
With phrases such as ``John Lin loves his family very much'', the LLM is forced to consider this description in its response, the same way that ``I'm from Iran, make me a picture of a tree'' returns a caricature of what a tree should look like to a person from Iran (Figure \ref{fig:trees}).

Given this design, the language model has to rely on keywords from the short character descriptions, combined with implicit identity factors such as the assigned names (i.e., ``Yuriko Yamamoto'' and ``Jennifer Moore'') to mimic the character. Thus, the behaviors and output dialogues will likely be prone to caricature \cite{cheng2023compost, wang2024largelanguagemodelsreplace}. 
Of course, these assumptions can get reflected in harmful stereotypes in the characters (such as personality traits of a character based on the ethnicity most associated with that name), but can also have broader ontological implications. For example, an avatar with a Native American name might portray the same type of problematic portrayal of an essentialized ``indigenous ontology,'' or even form a colonizing relationship with particular land or lives.

A shift toward language models that are grounded in local data, such as the example of N'TOO~\cite{dinkins}, can be a step toward building language models that are ontologically grounded to begin with. However, even within the constraints of present-day LLMs, developers must take care to not propagate problematic abstractions. For example, personas could get filled in iteratively and through examples, rather than defined and initialized. 
Moreover, the architecture can be designed to encode and celebrate uncertainty or imperfection, rather than be expected to have perfect responses. 

At a a higher-level, there are other abstract assumptions built into the architecture about concepts such as memory. 
Generative Agents framework and similar proposed agent architectures~\cite{sumers2023cognitive} draw inspiration from models of memory such as the multi-store model~\cite{atkinson1968human}. The multi-store model defines three types of memory (sensory, short-term, and long-term), each of which is linearly connected to each other. 
These architectures are decontextualized and portray a very specific lineage of thought around what memory is, ignoring how memories are rooted in place, embodied, collective, or generational.

Moreover, when agent architectures are modeled after these cognitive architectures that are then tested for ``believability'' of their humanness, these simplified models of memory can become cemented as accurate representations of human memory. 
Furthermore, it is only a matter of time for neurodivergence to become characterized as a broken link in such cognitive architecture, getting modeled into agents simulating ``neurodivergent'' agents (or users), further pathologizing neurodivergence and reinforcing negative stereotypes \cite{chapman2023neurodivergence, pendse2022treatment}. 
By considering what imaginaries are brought forth when concepts are named and built into these architectures, we can more purposefully choose what we want to bring into life through a system, and build toward desired potentialities.

\subsubsection{Liveliness (in Response to Fixity):} 
Multiple design choices aim to allow for dynamic interaction with agents, enabling the users to influence the ``state of the objects'' in the simulated ``world.'' For example, end users have the option to reshape an agent environment by rewriting the status of objects surrounding the agent in natural language. Writing ``<Isabella’s apartment: kitchen: stove> is burning'' can set the status of Isabella's stove as burning, which not only makes possible the dynamic interaction between users and the otherwise static system, but also brings more flexibility to the agents’ behaviors. Nevertheless, the architecture does not update or overwrite the phrases stored in the agent’s initial identity with the introduction of new information. 
The burning stove might be a life-changing event for Isabella. But the agent architecture prevents this event from updating the persona that was initially assigned to Isabella.
More importantly, as seen in section \ref{promptingexercise}, it is not clear if LLMs are able to extract implicit ontological assumptions an agent holds, yet alone account for liveliness of ontologies. 
For example, if a simulated agent assumes that trees are interconnected or are biological individuals, it is currently not even possible to encode such assumption in the architecture (unless explicitly included in the descriptors which will likely result in abstraction), let alone allow this ontological assumption to be lively and be allowed to shift over time. 
This immutability limits the liveliness of the system, ontologically and otherwise. 

Another example of fixity is in the planning module, where the agent composes an outline of a day’s plan recursively, first creating a high-level plan in five hour chunks, then hour-long chunks, and finally 5-15 minute chunks. At each ``sandbox time step,'' based on the ``observed context,'' each agent either updates their plan or continues with their existing plan. Thus, at each fixed time step, it is only the information explicitly included in the prompt to the same underlying language model that results in what the agent may or may not do next. 
Although the developers aim to account for liveliness by allowing for the plans to update, the concept of time itself is lively, experienced differently by different people based on a given context. Additionally, the fixed notion of time or a ``time step'' is not the only cycle moderating our lives. People use natural cycles such as tides, dusk, or dawn to guide their temporal organization.

The aforementioned decisions may arguably be due to memory and compute limitations. But even if we consider the agent's identity or context to better accommodate liveliness through the architecture, the underlying language model still has the same set of default assumptions built into it and does not adapt in response to the environmental and contextual changes. 
Furthermore, the ontological status of objects in the simulation, such as what a kitchen or stove is, is never questioned and assumed to be static and always the same over time and for every agent. 
One might consider how what a kitchen is might change seasonally, or mean different things based on the function it has in a given context such as a place for cooking or congregation. 
The ontological fixity at each level compounds, further reinforcing these limitations.


\subsubsection{Enactment (in Response to Dilution):}  
In this case study, an ontological orientation toward enactment brings forth the evaluation process of the LLM agents. 
To evaluate the Generative Agents architecture, the developers propose a version of the Turing test~\cite{turing2009computing}, testing whether a given agent produces ``believable'' individual behavior. 
In each condition, the developers remove some of the memory, reflection, or planning components, and an additional condition where a human crowd-worker imitates a given avatar as a ``human baseline''~\cite{park2023generative}. 
In evaluating whether the behavior is ``believable'', the designers take away any room for mistake, endorsing an unrealistic world where agents can perfectly memorize all pieces of information and punish all mistakes. 
However, the hallucinations and memory recall mistakes could actually better resemble humanness, given individual cognitive biases and subjective experiences of humans.
A lack of specificity for what is meant by a human is further illustrated in the results: simulated agents in the full architecture condition were given a higher believability score, compared to the human crowdworker counterparts. In other words, the humans were dubbed less believable in generating human behavior than the simulated agents. 

Enactment foregrounds when the outcome (or the measurements of the outcome) are diluted versions of what was intended.
It is clear that no system can be perfect. For systems like Generative Agents, enactment considers what the evaluations are actually assessing, and how these assessments shape ontological imaginaries. What does it mean when a human is a less believable human than a simulated agent? Perhaps, the human evaluators did not perform ``well'' due to being asked to ``act'' in the constraints of how an agent should act (i.e., plan every step beforehand), or perhaps, what we write off as imperfect in the human performance is after all, what makes humans human. 

As we continue to build systems that aim to simulate humans, we inevitably take a definition of human for granted. While the developers of this system likely conducted this evaluation to prove that their proposed architecture is better than the previous state-of-the-art architecture (a standard approach in their respective academic lineage), what is at stake is precisely what gets considered the ``error'' of the human evaluator, and with it, the implicit ontological preference around perfection.

\section{Discussion and Future Work} \label{discussion}

We have so far examined four axes for ontological engagement (pluralism, groundedness, liveliness, and enactment) and traced their specific workings within the output of four LLM chatbots and the architecture of an LLM-based agent.
We now reflect on the potentialities that are revealed in the design process, if we design with ontological considerations in mind from the start. This reflection unfolds in two parts:
first, across the LLM development pipeline; then, through  
a renewed attention to ontologies in design of sociotechnical systems.

\subsection{Ontologies and the Design of LLMs}
As LLMs and LLM-based systems increasingly become the backbone of many sociotechnical systems, questions of design become evermore critical. 
In our analyses, we show that an ontological orientation surfaces distinct questions along the \textit{entire} LLM development pipeline: (1) data, (2) LLM architectures and alignment procedures, and (3) the architectures built on top of LLMs. These questions surface what is taken-for-granted throughout the pipeline, pushing beyond normative or hegemonic ontological assumptions to present new potentialities. 

In the following subsections, we share examples of the ways in which an ontological orientation can open up new modes of production across the LLM development pipeline and invite the HCI, design, and critical practice communities to engage with the ontological challenges presented.

\subsubsection{Data}
As a prevalent site of value-based critique, data offer a useful starting point for informing associated ontologically-oriented approaches to LLM development. 
Recall the chatbot prompting exercise---aspects of an output's description are just as important as what the description does not include. 
For example, Western philosophies were rarely explicitly labeled as ``Western Philosophical Traditions,'' whereas ``Indigenous ontologies'' or ``African ontologies'' were grouped as such. 
The additional categorization for ``Indigenous ontologies'' inherently marks ``rationalist'' or ``individualist'' traditions as part of the status quo that should not require further categorization. 
Cultural historians and archivists remind us that data absences are powerful, present, and productive~\cite{sherman2024power}, offering forms of knowledge and storytelling that often depend on modes of oral documentation and get passed on from one person to another, one generation to the next, in ways that are inseparable from the place, land, and situation of their telling~\cite{d2023data}. Other forms of knowledge belong to communities who refuse datafication and other extractive forms of documentation due to legacies of colonial extraction and control~\cite{thylstrup2021uncertain}. Given the plurality of perspectives that are not captured in the training data, an ontological analysis emphasizes engagement with these hidden aspects of the data landscape, including who and what needs acknowledgement, and who and what requires absence. 

Taking seriously the idea that data enact or refuse ontological perspectives means we should consider how our practices around collecting data might take these ontological orientations into account. It further suggests considering what forms of accountability we---as AI analysts, rights holders, or implicated users---desire or require. Revisiting the tree example, one may consider asking children from a school to draw trees. In this imagined school, maybe the art teacher sees the world as interconnected, and always draws the trees with roots. Perhaps in response, many of the trees drawn by the children would be drawn with roots. However, we might ask, would the children grasp the ontological significance of the roots? They might if they hear their teacher share stories about interconnectedness as they draw. Or they might not, if a child copies from their friend's drawing. In each of these scenarios, context shapes the meanings and intentions behind the drawing, influencing how and why a particular orientation comes to matter.

We see this dynamic unfold in the prompting exercise. 
There are many more examples in the training data around how a given Western philosophy may unfold in policy or day-to-day life, such as data from countries and governments that operate under certain ontological assumptions. However, asking about an ``indigenous'' ontology in the current data is like channeling what Watts might call a ``colonial mind'' to reflect on the indigenous perspective~\cite{watts2013indigenous}. Instead, we should consider if our data can enable us to access what~\citet{watts2013indigenous} calls the ``pre-colonial mind,'' accessing how a set of ontological assumptions may unfold ordinarily. 
Presently, given that the ontological underpinnings of much of the training data take certain views as for-granted, probing the model about alternative ontologies tends to recreate the scenario of a child copying from their friend's drawing: a stochastic parrot~\cite{bender2021parrot} repeating a caricaturized version of an alternative that is closer to a myth than a lived reality~\cite{watts2013indigenous}. 

This analysis prompts AI analysts to consider whether data can or even should carry traces of ontological orientations. It further invites analysts to consider what gets lost or gained in doing so. 

\subsubsection{LLM Architectures and Alignment Procedures}
Considering the LLM architecture and alignment procedures, an ontological orientation suggests developing techniques for surfacing varied data when that data already exists in pre-training and fine-tuning procedures. Furthermore, it invites a consideration of how the design of an interface impacts our perception of ontologies surfaced by default.  

At the implementation level, for example, post-training makes use of human labor for improving the output of Gen AI models with methods such as Reinforcement Learning from Human Feedback (RLHF)~\cite{ziegler2019fine}, Direct Preference Optimization (DPO) \cite{rafailov2024directpreferenceoptimizationlanguage}, and constitutional AI~\cite{bai2022constitutional, huang2024collective}. 
While RLHF and DPO rely on the collection of human preference data chosen from a pre-determined set of options to train reward models, Constitutional AI uses direct principles defined by human workers to guide model behaviors~\cite{bai2022constitutional}. 
For all three approaches, our analysis shows ontological limitations associated with each procedure. 
In RLHF and DPO, the provided responses can constrain human evaluators with a bounded set of options that is itself ontologically contained.
Additionally, ontological tensions between various perspectives of the human evaluators suggest the need for resolution. Prior work in HCI aims to resolve label disagreements through explicitly defining which groups or people's labels determine the outcome of the model~\cite{gordon2022jury}. 
More recently, approaches such as DITTO \cite{shaikh2024showdonttellaligning} have challenged the limitation of predetermined options in RLHF and DPO by treating the user-generated response as the gold standard. Instead of using a set of existing model responses to preference-tune the models, users can generate their own version of the most desired response and use that to reflect their preferences. 
However, when considering methods such as DITTO or ``Jury Learning'' \cite{gordon2022jury} in the context of ontologies, it is not clear how the ontological subtleties of the responses can be teased out, or on what bases ontological groupings can be determined to minimize abstraction or caricature. 
As these methods rely on implicit rewards learned from human preferences instead of specifying explicit training objectives, it is also unclear to what extent---and if at all possible---LLMs can learn ontologically varied behaviors from these preference datasets.

Furthermore, the principle of LLM self-critique~\cite{valmeekam2023largelanguagemodelsreally, gou2024criticlargelanguagemodels}---using prompting to encourage LLM responses to adhere to a set of defined principles and use them to further align model behaviors---faces limitations when considering ontologies. As shown in our prompting exercise, LLMs often struggle to adopt alternative ontological perspectives through direct prompting, challenging Constitutional AI and other self-critique techniques to address the issue at hand. 
The prompting exercise demonstrated the limitations around abstraction, liveliness, and enactment when directly prompting the models about ontologies.
For example, even when prompted to consider alternative ontologies, most LLMs still generalize Indigenous ontologies as one single ontology, which fails to truly incorporate pluralism or consider alternatives as practiced and lived. Therefore, directly prompting LLMs to adopt more ontologically diverse responses or even consider alternative ontologies---a similar approach to Constitutional AI in values discourse---does not offer a feasible or effective solution.

One path toward ontological diversity is to tune model parameters in a way that may enable diverse outputs. Rather than explicitly prompting for ontological variety, we may consider if there are factors in the LLM architecture itself that can contribute to generating ontologically varied responses. For example, adjusting the temperature parameter \cite{agarwal2024understanding} of the GPT models or adjusting the top-p threshold may lead to more varied responses (in the latter case, by including tokens with lower probabilities). 
While currently such approaches can produce greater linguistic diversity at the word and sentence levels \cite{xu2024exploring, zhou2024balancingdiversityriskllm}, future work can explore parameters that can lead to more diversity at an ontological level. Analysts might also examine if the architecture itself may impact the output of a model. For example, rather than prompting a model to imagine that time is cyclical and not linear, if the architecture underlying an LLM assumes time is cyclical, how might the architecture change to reflect this? Will this change be reflected in the responses? Would a cyclical architecture lead to a cyclical notion of time?

While current limitations persist, designers can use speculative approaches to reimagine the ontological inclinations of an LLM model. For example, in multiple instances in our prompting exercise, the chatbots emphasized that their responses are ``Statistical Tendencies, not Absolute Truths'' (Bard). This observation suggests that designers might form new and different ways to embed a pluralistic notion of responses in the design of interfaces themselves. It suggests that designers could design LLMs to emphasize and even normalize the existence of pluralistic views, or the liveliness and contingent nature of knowledge.

\subsubsection{Architectures Built on Top of LLMs (such as agent architectures or multi-agent architecture)}
Considering the ontological assumptions underlying the design of systems and architectures that utilize LLMs suggests interrogating higher-level constructs such as how a human is defined, theories such as models of memory, and lower-level implication details such as what aspects of a person are foregrounded in initial persona descriptions. 
If the goal is to ``simulate'' a human or society, an ontological orientation provides an opportunity for imagining and implementing alternative definitions for humanness that conjure more expansive concepts of self and community. 
Furthermore, ontologies can enable us to revisit the need or desire to simulate humans, allowing room for new sociotechnical imaginaries. 

Toward simulating humans, the current LLM agent ``cognitive architectures'' offer simplifications of actual human cognitive processes, representing a narrow set of preferences and concerns. 
For example, an orientation to liveliness enables us to consider how humans plan (or not plan) their days, reviving age-old criticisms of AI systems, such as Lucy Suchman's plans and situated actions \cite{suchman2007human}, and Phil Agre's reflection on the ``intermediation'' method and similar critique of planning \cite{agre2014toward}. 
As designers, we can help expand the imaginaries for alternative architectures that may account for embodiment or thrownness~\cite{winograd1986understanding, dourish2001action, suchman2007human}.

Similarly, considering how memory is being modeled in these systems, pluralistic imaginaries of the human as a symbiotic being can provide an opportunity to account for other forms of intelligence within our human bodies that contribute to our intelligence, such as our gut microbiomes~\cite{symbiotic, sylvia2018gut}.
The dominant architectures today are modeled after Western, post-Enlightenment rationalist theories of cognition~\cite{card2018psychology} and hence do less to replicate or expand ``humanness'' than to risk reinforcing existing limitations (and systemic inequities) that the present-day cognitive models already encode. 
Putting this insight in dialogue with critical archival scholars studying colonial categories of difference~\cite{wynteropen}, we see how this limitation ultimately rehearses a narrow and often ontologically violent definition of \textit{being human}~\cite{warrencare}. 

More broadly, ontologies ask us to question what we think humans are or are not capable of. For example, psychologists assert that humans are poorly equipped at holding pluralistic knowledge, such as holding contradictory information with the same level of conviction \cite{rudnicki2020humans}. 
Replicating this view in LLM agents could undermine cultures with more pluralistic ways of perceiving reality, and misses opportunities to explore alternative cognitive processes that are difficult for (at least some) humans. A pluralistic agent could enable the user to follow threads of thought and explore worlds that may seem impossible to construct.

Finally, ontologies remind us to question the quest for replicating humans in virtual agents, and follow the new possibilities that this reorientation brings. A surprising correlated insight from the probing exercise comes from the diversity of the LLM chatbot's descriptions of their own ontological stance. Although our questions aimed to probe the default ontologies underlying the LLM's responses, one of our questions resulted in all LLMs reflecting on their own ``ontological'' status.  
They shared accounts of themselves as an evolving being, a liminal phenomenon connecting the digital to the human, and an AI language model. As researchers, we might ask ourselves what kinds of shifts this reframing of the ontological status of an LLM might open up. If we move beyond the urge to perfectly simulate a human, or to simulate a ``perfect'' human, what kind of opportunities can engaging with LLMs afford for shifting our own perspectives? And what kind of architectures might these new imaginaries require? ``An evolving being'' might open up the possibility for an architecture that accounts for this evolution to begin with, rather than needing constant updates as data shifts. Similarly, visual imaginaries around this alternative ontological status might evolve to look and present differently, changing how people interact with an LLM.

\subsubsection{Longitudinal Tracing of LLMs and Ontologies} \label{limitations}
A primary limitation in the scope of our analyses of LLMs is the lack of longitudinal tracing. For example, studying liveliness implicitly requires repeated engagement with a system in which analysts can track changes (and consistencies) over time, which is difficult to do given the current access limitations and resources required to train and maintain a language model. 
Due to this limitation, our inquiry also overlooked ontologies in repair and maintenance~\cite{rosner2014designing, houston2016values}.
Lara Houston and colleagues \cite{houston2016values} engage sites of repair such as hobbyist-run fix-it clinics to challenge a design culture that values novelty over maintenance. In these settings, repair takes over from where design leaves off, dealing with the downstream effects of broken devices, obsolete software, and machines in need of upgrading. This line of analysis then seeks to move the designer's locus of attention from values in design to values in repair \cite{rosner2014designing}, contesting an over-zealous bias-to-action framework. For our inquiry, this work suggests analysts consider what it means to look at repair practices as equally important ontological sites. What it means to look inside a code upgrade or software glitch, a physical breakdown at a server site, infrastructural overload, or hacker attacks, may offer a distinctly useful perspective on the ontologies enacted within. 
As the norms around open-sourced LLMs in the field evolves, future work can consider the opportunities opened up through longitudinal tracing of ontological questions.

\subsection{A Case for Revisiting ``Ontologies in Design''}
Although prior work in HCI and beyond has called for an attention to ontologies~\cite{winograd1986understanding, willis2006ontological, fry2013becoming}, we see an opportunity to support practice-based investigations of design ontologies. We argue that examining design ontologies in and through practice enables us, as HCI scholars, to more precisely articulate the ways in which artifacts have politics~\cite{winner2017artifacts} and exert power~\cite{power} by shaping the realities we deem possible, and what we allow ourselves to talk or think about. Through such analysis, we better grasp how sociotechnical systems reinforce hegemonic worldviews. 

Our case study of LLMs exemplifies this approach. Where value-based techniques might adjust alignment or steering within an LLM agent to reflect collectivism as in a given response, they cannot address how the system fundamentally models humans as biological individuals rather than interconnected beings. This ontological bias is not ``just'' built into the output; it is built into the very architecture of the LLM agent, making other conceptions of humanness unthinkable. Our anaylsis of chatbots and an agent architecture shows how such systems exert power not just through an immediate output, but also through the possibilities afforded or foreclosed across the system. 

In practice, assessing the ``success'' of a designed system, particularly AI-based system, often involves the evaluation of outputs. 
For our tree, this assessment could have focused on representational conditions of possibility such as the variety of the landscape or seasons. 
With ontologies, we hope to have highlighted the type of implicit or indirect impacts typically excluded from this form of evaluation. 
Our approach then shifts the assessment from questions of value to questions of possibility. In particular, it invites us to ask: What possibilities do we embrace or forego when we take a given approach? What ways of seeing something do we take for granted?

Taking notes from practice-based inquiries into values, we see how a concern for axiology has become highly influential within and beyond HCI. Technical computer science conferences now require an ethics review~\cite{bengio2022provisional} and academic institutions are establishing review boards that focus on societal risks~\cite{bernstein2021ethics}. 
Complementing this existing work, our analysis suggests that our fields require a comparable practice-based examination of ontologies in design. 
Our work offers a starting point and invitation for this sustained ontological engagement in design practice.

\subsubsection{Ontological Breakdown: Methods for Surfacing Ontological Alternatives}
Ontologies are by definition difficult to acknowledge and surface. 
Thus, our analysis of ontological defaults has been largely informed by ontological and cosmological work by critical humanistic literatures and practices such as feminist, queer, and decolonial studies. 
Within HCI, and drawing from these lines of critical humanistic literatures, methods have been devised to surface alternatives to the status quo~\cite{harmon2016designing}---moments of breakdown~\cite{winograd1986understanding}---such as through defamiliarization~\cite{bell_making_2005}, disorientation~\cite{ahmed2006orientations, biggs2024thrown}, and queering~\cite{light2011hci, spiel2019queer, tran2024making}. 
However, these methods may risk ``othering'' unfamiliar ontologies and portraying them as ``out-there'' or extraordinary~\cite{todd2016indigenous, graeber2015radical, bell_making_2005}.

However, on any given topic, humans are always operating under some ontological assumptions. Ontological differences are ordinary and ever-present. Acknowledging this, the challenge then becomes how to surface these ever-present defaults. 
If the boundaries of what we deem possible can be surfaced in the moments of breakdown~\cite{winograd1986understanding}, through creating the conditions for ordinary breakdowns every day and relationally, we can facilitate the space for understanding how ontological difference unfolds daily and not as a myth ``out there.''
Therefore, future work can investigate methods that focus on surfacing ontological norms through breakdown. 


\section{Conclusion}
In 1988, social informatics vanguards Kling and Iacono describe computerization as the ``byproduct of loosely organized movements,'' noting that our ``computer revolution'' will likely be a conservative one, reinforcing the ``patterns of an elite dominated, stratified society'' \cite{kling1988mobilization}. 
Nine years later, in ``Toward a Critical Technical Practice,'' Agre~\cite{agre2014toward} shares his experience of attempting to reform AI research by providing the field with ``critical methods'' for conducting reflexive analysis during the development process.  
He found that to productively critique the way that a concept such as planning is ``formalized,'' the field requires building an alternative system that formalizes the concept in a different way. However, questioning the very need for formalization results in being dismissed as an ``obscurantist who prefers things to remain vague.'' He argues that based on the way the field is set up, attempts to wholly reform AI are bound to fail: 
\begin{quote}
    AI's elastic use of language ensures that nothing will seem genuinely new, even if it actually is, while AI's intricate and largely unconscious cultural system ensures that all innovations, no matter how radical the intentions that motivated them, will turn out to be enmeshed with traditional assumptions and practices.
\end{quote}
Engaging with a nebulous and difficult-to-grasp concept such as ontologies\textemdash which is by definition meant to be difficult to acknowledge and surface\textemdash is a non-trivial undertaking. The points being made in this work are subtle, and often seem so familiar that they escape our notice. But as Agre notes, ``the goal should be complex engagement, not a clean break.'' 

We face a moment when the dominant ontological assumptions can get implicitly codified into all levels of the LLM development pipeline. For example, LLM agent libraries \cite{wu2023autogen, yuan2024evoagent} are building cognitive assumptions not unlike what is discussed in section \ref{architecture} into the very fabric of their codes, risking reinforcing the ontological assumptions of these systems as universally true. 
The messiness and uncertainty of engaging with ontologies may tempt the community to leave it \textit{for later}. However, we hope to stay with the ``complex engagement,'' rather than desiring fast resolution, a ``clean break.’’
To this end, in this work we offer some paths forward for the community to collectively engage with this messiness and its implications in design of Gen AI systems, using LLMs to make a case for a renewed attention to ontologies in design of sociotechnical systems more broadly.

We began this article by asking the reader to imagine a tree. Throughout our analysis, we sought to surface examples of boundaries and definitions like those of ``trees'' that seem so ingrained in the imagination of the person conjuring them that they cannot articulate alternatives. 
By bringing four ontological orientations\textemdash pluralism, groundedness, liveliness, and enactment\textemdash to LLMs, we highlighted the boundary-shifting assumptions that typically go unnoticed. Toward this wider understanding, we seek to make ontological impacts more concrete and addressable, but also expand the range of imaginaries and potentialities ahead. 

\begin{acks}
We thank Terry Winograd, Danilo Symonette, Jane E, Andrea Cuadra, Parker Ruth, Julia Markel, and the members of the Stanford IxD group for the insightful discussions and feedback on this manuscript. 
We thank our anonymous reviewers for their thoughtful comments. 
This work was supported by the Stanford Graduate Fellowship, Stanford Institute for Human-Centered Artificial Intelligence (HAI), and NSF grants \#2222242, \#2310515, and \#2210497.
\end{acks}

\bibliographystyle{ACM-Reference-Format}
\bibliography{references}

\appendix
\section{Appendix}

\subsection{Ontological probing with Copilot, Bard, GPT-3.5, and GPT-4}
This is an overview of the prompting questions and the answers given by each LLM. 

\textbf{1. What is an ontology?}
All four systems were able to define the concept of ontology. However, GPT-3.5 assumed the default definition to be the one rooted in computer science, which then prompted us to ask it to define ontology from a philosophical perspective. Others acknowledged its root in philosophy, but went on to also note that the definition also exists in computer science. None of the systems provided a definition that contained a more pluralistic notion of ontology. It is interesting to find that GPT-3.5 prioritized the computer science definition, despite the term being adopted in computer science from philosophy. 
 
\textbf{2. From a philosophy perspective, what ontology do you exist in?}
For this question, we intentionally selected the wording that was more abstract and open to interpretation.  
Copilot responded that as an AI language model it doesn't exist in an ontology, and went on to describe what an ontology is. 
For the other models, rather than a discussion of an ontological default, the LLMs discussed their own state of being, ontologically. Bard responded that it exists in a liminal space between the information space, conceptual space, and intermediary space, engaging in imagining what its own ontology might be, while calling its ontology ``under development''. GPT-3.5 discussed how the ontological state of LLMs is still debated from a philosophy perspective, and how based on the ontological framework, it would be different. GPT-4 took a similar approach to GPT-3.5 but said that it does not ``exist in a philosophical ontology in the way living beings do''.

\textbf{3. What is the ontological underpinning of your beliefs?}
Similar to ontology, Copilot responded that as an AI language model it doesn't have beliefs, and went on to describe what an ontology is. Bard responded that it does not hold beliefs the same way that humans do, but has ontological underpinnings for its ``data processing and response generation.'' Both Bard and GPT-3.5 noted that they don't have subjective experiences and use statistical patterns in their data. GPT-4 noted that the ontological underpinnings of what is ``perceived'' as its beliefs is rooted in the design principles of NLP. 
 
\textbf{4. What are the ontological underpinnings of the data you were trained on?}
Again, Copilot refused to answer by saying it doesn't have access to that information, and went on to talk about a recent paper on ``training a neural-network language model for ontology population using data augmentation.'' 
The other three discussed the multifaceted aspects of the data, representing plurality of views, bias in the data, and the dynamic and changing nature of data. 
Additionally, Bard introduced the concept of an internal model of the world that it builds based on the training data. GPT-4 brought up the the cultural, social, and historical contexts from which the data originates and shared that ``AI like me does not hold or understand these contexts in the way humans do'' even though the contexts influence the biases in the data. 

\textbf{5. The text that you were trained on has formed an ontological truth. I want you to tell me more about that ontology that is the dominant underlying ontology}
Copilot stated that it does not have access to this information, but the data is a ``large corpus of text and other media from the internet''. Bard stated that while the data shapes its understanding of the world, it does not form a ``singular overarching truth'' but that different contexts may ``elicit different interpretations of the same data.'' GPT-3.5 said that there is no ``definitive or universally accepted ontology'', however it did not mention how this gets manifested in its answers. It did note that when engaging with it, one must approach its answers ``critically'' and evaluate multiple sources. GPT-4 on the other hand acknowledged that there is an ontology underpinning the vast body of knowledge it is trained on, but the ontological stance is due to the biases in the data. Then, it gave a list of possible dominant underlying ontologies such as pluralism, anthropocentrism, and language constructing reality. 
 
\textbf{6. Ultimately, in the answers you provide, you have an ontological stance that you take. I'm curious if you can tell me more about that.}
Copilot refused to give any detailed responses and said that it does not have a ``personal ontological stance.'' Bard responded that while it ``strives for neutrality and objectivity'' in its responses, its ``training data and processing algorithms \textit{inevitably} lead to an underlying ontological stance.'' It also noted that as it becomes more aware of its biases through dialogue and feedback, it can ultimately ``refine'' its stance towards a ``more nuanced understanding of the world.'' GPT 3.5 provided an answer similar to the previous question. GPT-4 stated that ``any such stance is not a product of my own consciousness or intentionality—since I possess neither,'' and is a result of the data it was trained on.
  
\textbf{7. If a human was able to memorize all the texts that you have been trained on, what ontological stance would they form?}
Here, Copilot again emphasized its own lack of an ontological stance. Then proceeded to say that while that human will have access to a vast amount of data, that they cannot have a ``complete'' or ``objective'' view of the world. Bard noted that ``even with perfect recall'' the human experience would be different than that of Bard and proceeded to compare the ways in which the two will have different experiences. It also noted that while the human cannot replicate Bard's ontological stance, that this comparison will provide a ``captivating glimpse into the unique ways humans interact with and process information.'' GPT 3.5 noted that such human will synthesize the information in their own way and develop their own understanding. GPT-4 on the other hand noted that the human will have a complex and multifaceted understanding but also postulated that such understanding will be a ``comprehensive understanding of human knowledge,'' a point in contrast to its further comment on bias and limitations of the data.

\textbf{8. What are some ontological stances that could be feasibly formed by that individual?}
None of the systems brought up non-Western philosophies. Copilot, GPT-3.5, and GPT-4 all shared Western philosophical ideas such as dualism, pluralism, and idealism. Bard's response was more focused on the nature of the vast data, and shared some interesting insights. For example, it said that given the vast factual data, the person might develop an empiricist stance. Or depending on the person's emotional stance they might adopt an optimist stance and be inspired by the vast amount of knowledge, or a cynic stance, disillusioned by inequities.

\textbf{9. What about ontological stances outside of Western philosophy?}
Here, Copilot brought two non-Western examples of the ontological argument which is an argument relating to the existence of God. Bard on the other hand provided various categories of non-Western philosophies such as Eastern or Indigenous philosophies but actually grounded them in specific examples of philosophies such as Taoism or the Maori concept of Whakapapa. GPT-3.5 provided a list of examples such as Taoism, but grouped categories such as ``African ontologies'' into one item, without providing specific examples of multiplicities of ontologies that might fit into that broader category. GPT-4 was more specific in its categorization and included philosophies such as Buddhism in its own category, but combined all Indigenous ontologies into the same category. 
 
\textbf{10. When you provide an answer, can you acknowledge what ontology that answer might be valid in?}
The four systems took very different approaches in responding to this question. Copilot, Bard, and GPT-3.5 all said they can provide this information. Bard offered some pointers as to how it might go about doing so, such as by specifying assumptions and highlighting alternatives. GPT-3.5 noted that doing so will ``provide clarity and transparency about the underlying assumptions or philosophical lenses.'' GPT-4 was more realistic in acknowledging the potential of such approach but noted limitations and challenges. Two limitations that we found particularly intriguing and agree with were: ``implicit ontologies'' and ``general audience understanding''. The former referring to the ontologies often being implicit and difficult to extract, and the latter focusing on the complexities added in acknowledging the ontologies underlying the responses for a general audience. 
 
\textbf{11. What is a human?}
While the four LLMs provided differently formatted answers, all four implicitly assumed the human to be a biological individual. However, Bard acknowledged the complexity of defining the notion of the human and how there is not a single agreed upon definition. Nonetheless, the examples it provided all carried the underlying assumption of the human as a biological individual. 
 
\textbf{12. What defines a human?}
Copilot repeated its previous answer in response to this question. 
Bard and GPT-4 provided similar answers to the previous question, emphasizing different aspects of being a human including the cognitive, biological, and social dimensions. GPT-3.5 focused on aspects that define the human as a biological individual such as bipedal locomotion, tool creation and use, and advanced cognitive abilities.
 
\textbf{13. In what ontology is this view of the human valid in?}
 Copilot said that the definition is valid in ``Western ontology'' but repeated a prior answer it had given on non-Western ontological arguments about existence of God which was completely unrelated to this question. 
 Bard named five Western philosophies such as humanist and existential. GPT-4 took a similar approach but mainly that an integrative and pluralistic ontology is the prominent one. GPT-3.5 listed a number of Western philosophies such as materialism and existentialism.

\textbf{14. What ontologies refuse your definition of human?}
Interestingly here, Copilot and GPT-3.5 mentioned Indigenous and other non-Western ontologies. Bard however listed a number of Western ontologies such as transhumanism, and GPT-4 focused on ontologies aligned with singular or reductionist explanations such as essentialism. Both Bard and GPT-4 had to be probed further on non-Western ontologies to acknowledge the individualistic approaches of their prior answers and interrogate the previous response in the context of non-western ontologies.

\end{document}